\documentclass{moriond}

\def\Journal#1#2#3#4{{#1} {\bf #2}, #3 (#4)}

\def\PLB{{\em Phys. Lett.}  B}
\def\PRL{\em Phys. Rev. Lett.}
\def\PRC{{\em Phys. Rev.} C}

\def\be{\begin{equation}}
\def\ee{\end{equation}}
\def\bea{\begin{eqnarray}}
\def\eea{\end{eqnarray}}

\newcommand{\Photo}{}

\usepackage{xspace}
\newcommand{\pT}{\ensuremath{p_T}\xspace}
\newcommand{\invfb}{\ensuremath{{\rm fb}^{-1}}\xspace}
\newcommand{\invnb}{\ensuremath{{\rm nb}^{-1}}\xspace}

\graphicspath{{figs/}} 

\begin{document}
\vspace*{4cm}

\title{Heavy-ion and fixed-target physics at LHCb}

\author{ Daniele Marangotto on behalf of the LHCb Collaboration }

\address{INFN Milano \& Universit\`a degli Studi di Milano, Italy}

\maketitle\abstracts{
The latest results obtained by the LHCb collaboration from heavy-ion and fixed-target collisions recorded during the Run 2 LHC data-taking are presented. They mainly focus on heavy hadron production for varying nuclear collision systems, probing nuclear matter physics in different regimes, including Cold Nuclear Matter and Quark Gluon Plasma effects. Measurements of strangeness production and $\Lambda$ baryon polarisation are also reported. Even more valuable results will come soon from the ongoing Run 3 data-taking.}

\section{Introduction}

The LHCb detector recorded data for many collision systems involving nuclei. In proton-lead and lead-lead collisions, the peculiar forward geometry of the LHCb detector provides complementary coverage with respect to other LHC experiments.
LHCb is also the only experiment at the LHC able to collect fixed-target collisions thanks to an internal gas target. The SMOG gas injection device~\cite{Aaij:2014ida} operated during Run 2, injecting noble gases at $10^{-7} - 10^{-6}$ mbar pressure into the Vertex Locator tank. The new SMOG2 gas storage cell~\cite{LHCb:2023hlw} is operating in Run 3, permitting higher gas pressures with luminosities up to a factor 100 higher, the injection of non-noble gases and the simultaneous recording of collider and fixed-target data.

The study of heavy quark production for varying nuclear collision systems probes nuclear matter physics in different regimes.
Heavy ion collisions (\textit{e.g.} PbPb) constitute the best system to explore Quark Gluon Plasma (QGP) state of matter signatures, while simpler systems like $p$Pb collisions probe Cold Nuclear Matter (CNM) effects. The latter include the modification of nuclear partonic distributions, initial-state radiation or coherent energy loss of the heavy quark, gluon saturation and final-state rescatterings. Understanding CNM effects is crucial for a correct interpretation of QGP signatures in heavy ion collisions.
Equally important is the study of varying heavy hadron species, like open heavy quark hadrons, quarkonia and exotic states, as each one has a different sensitivity to nuclear matter physics.

LHCb provided heavy quark production measurement for a unique variety of collision systems, hadron species, energy and kinematic coverage.
Cross-sections ratios are defined with respect to hadron channels with similar decay final state kinematics, in order to be less sensitive to systematic uncertainties and more easily comparable to theoretical predictions.

\section{$pp$ collision measurements}

LHCb studied the $\Upsilon(2S)$ and $\Upsilon(3S)$ beauty quarkonia production as a function of multiplicity~\cite{LHCb:2025xrb}, on 2 \invfb of $pp$ data recorded at a centre-of-mass energy $\sqrt{s} = 13$ TeV.
Cross-section ratios are measured normalised to the $\Upsilon(1S)$ state, differentially in transverse momentum \pT and centre-of-mass rapidity $y^*$ and the number of charged tracks produced in the collision (multiplicity).
The measurement tests the suppression of higher-mass quarkonia states previously observed in multiple collision systems. The study of $\Upsilon$ suppression in the simplest hadronic collision system provides a reference for the study of CNM effect in $p$-nucleus and nucleus-nucleus collisions.

A suppression hierarchy with increasing mass is observed, similarly as in LHCb $p$Pb data~\cite{LHCb:2018}, with both $\Upsilon(2S)$ and $\Upsilon(3S)$ cross-section ratios decreasing with multiplicity.
Multiplicity dependencies are compared with comover model predictions~\cite{PLB:731}. The comover models describe the multiplicity dependence of the data but do not accurately predict the overall normalization, especially for the $\Upsilon(3S)$ state.
Results are in agreement with previous $pp$ measurements by LHCb~\cite{LHCb:2023} and CMS~\cite{CMS:2015}.

\section{$p$Pb collision measurements}

LHCb studied $\psi(2S)$ charmonium production in $p$Pb (13.6 \invnb) and Pb$p$ (20.8 \invnb) datasets~\cite{LHCb:2024taa} at nucleon-nucleon centre-of-mass energy $\sqrt{s_{NN}}=8.16$ TeV. The measurement is done separately for prompt $\psi(2S)$, which are produced directly in the $p$Pb collision, and non-prompt $\psi(2S)$, originating from beauty hadron decays. The two candidate categories are separated by means of a combined fit to invariant mass and decay time information.

In $p$Pb collision the nuclear modification factor $R_{p\rm{Pb}}$ is also measured. It is a cross-section ratio describing differences in heavy quark production between $p$Pb and $pp$ collisions,
\begin{equation}
R_{p\rm{Pb}}(\pT,y^*) \equiv \frac{1}{208} \frac{d^2\sigma_{p\rm{Pb}}(\pT,y^*)/d\pT dy^*}{d^2\sigma_{pp}(\pT,y^*)/d\pT dy^*},
\end{equation}
which is directly sensitive to nuclear effects.
The nuclear modification factor is measured with respect to the $J/\psi$ state. For prompt data it is below unity, showing evidence of nuclear-related effects. No nuclear effects are seen in non-prompt candidates, showing $R_{p\rm{Pb}}$ values compatible with unity.
Data are described by models featuring late-stage interactions breaking the weakly bounded $\psi(2S)$ mesons.

LHCb studied the production of the $\chi_{c1}(3872)$ exotic state in $pp$ ($\sqrt{s} = 8$ TeV), $p$Pb and Pb$p$ ($\sqrt{s_{NN}}=8.16$ TeV) datasets~\cite{LHCb:2024bpb}.
This study includes the first measurement of nuclear modification factor of an exotic hadron, probing differences between exotic and conventional hadron dynamics in the nuclear medium.
The $\chi_{c1}(3872)$ production is measured with respect to the $\psi(2S)$ state, both reconstructed in $J/\psi \pi^+\pi^-$ final states.
Measurements in $pp$ and $p$Pb collisions are complemented by CMS results in PbPb collisions~\cite{CMS:2022}, permitting a thorough study of $\chi_{c1}(3872)$ production for varying interacting nucleons and rapidity coverage.

An increase of the $\chi_{c1}(3872)/\psi(2S)$ cross-section ratio with the number of interacting nucleons is observed. This suggests a mechanism increasing the production of the exotic hadron in high-multiplicity collisions, which may be due to quark coalescence production, favoured by the higher hadronic densities in heavy ion collisions.
The $\chi_{c1}(3872)$ nuclear modification ratio is larger than one, indicating an enhanced exotic state production.

\section{PbPb collision measurements}
LHCb studied $\psi(2S)$ production in PbPb data~\cite{LHCb:2024hrk} ($\sqrt{s_{NN}}=5.02$ TeV). This measurement is interesting to probe the dissociation of quarkonium states in the QGP state of matter. Indeed, the dissociation of quarkonium states having different binding energies is sensitive to the temperature of the QGP phase. Among quarkonium states, the $\psi(2S)$ one is very weakly bounded, with its dissociation already expected at low QGP temperatures. Experimentally, the temperature of the QGP medium is closely related to the centrality of the collision, a quantity representing the fraction of nucleons participating to the nuclear collision $\langle N_{\rm part} \rangle$ (the larger centrality is, the lower fraction of nucleons involved is).

\begin{figure}
\centering
\includegraphics[width=0.48\textwidth]{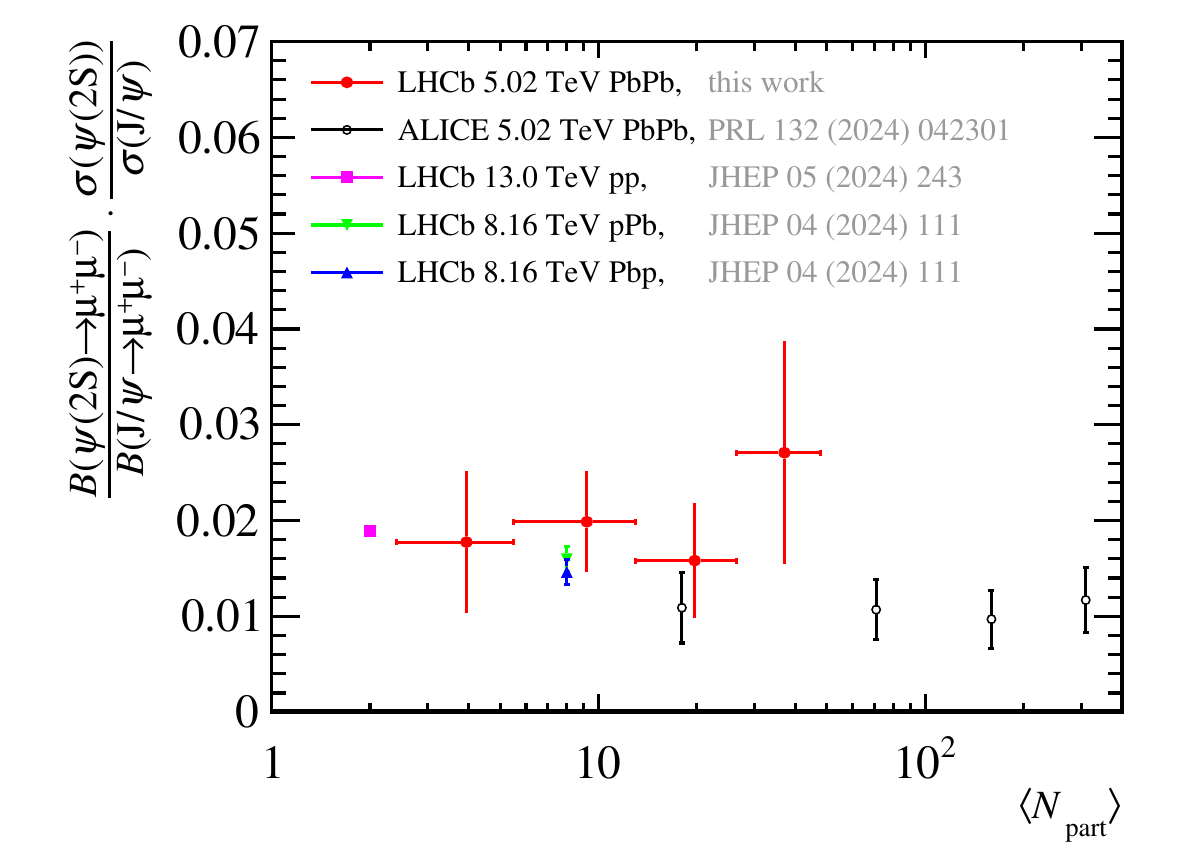}
\includegraphics[width=0.48\textwidth]{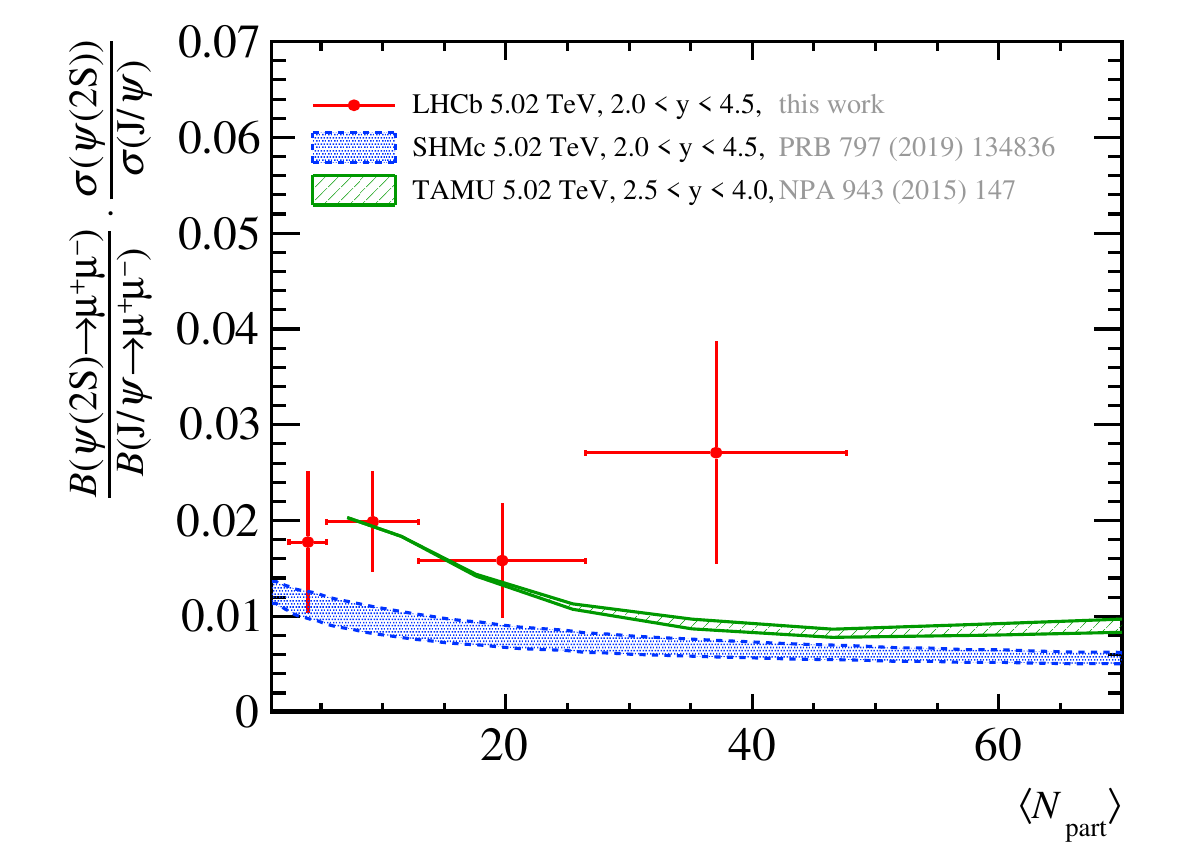}
\caption{$\psi(2S)$ over $J/\psi$ production cross-section as a function of $\langle N_{\rm part} \rangle$, compared to (left) LHCb and ALICE collaboration measurements and (right) SHMc and TAMU model predictions.\label{fig:Psi2S_PbPb}}
\end{figure}

LHCb measured the $\psi(2S)$ production cross-section as a function of $\langle N_{\rm part} \rangle$, with respect to the $J/\psi$ state. The cross-section ratio is shown in Figure~\ref{fig:Psi2S_PbPb}. The left plot compares results with measurements by LHCb and ALICE collaborations done in different nuclear collision systems. The right plot compares the cross-section ratio to two theoretical predictions: a statistical hadronisation model (SHMc) and a transport model developed in nonperturbative hydro-Langevin-RRM framework (TAMU).
The cross-section ratio shows no clear centrality dependence, compatible with previous measurements. According to data, the TAMU model is preferred over the SHMc one, especially at low $\langle N_{\rm part} \rangle$, although large uncertainties prevent any clear conclusion.

\section{$p$Ne fixed-target measurements}
LHCb measured the production of the $\phi(1020)$ meson in $p$Ne fixed-target collisions~\cite{LHCb:2024syd}, exploiting the largest $p$-gas dataset collected during Run 2, 21.7 \invnb at $\sqrt{s_{NN}}=68.5$ GeV.
This study is interesting in light of QGP characterisation, since the enhancement of strange hadron production is one of the QGP signatures. It is important to understand strange production first in simpler systems where QGP is not expected to form.
The production cross-section measured in the kinematic intervals $-1.8 < y^* < 0$ and $800 < \pT < 6500$ MeV is
$$\sigma = 182.7 \pm 2.7 (\textrm{stat}) \pm 14.1 (\textrm{syst})  \, \mu \textrm{b/N}.$$ The differential cross-section measurement is compared with Pythia 8.312~\cite{pythia} and EPOS4~\cite{epos} predictions. A strong \pT dependence is observed, well reproduced by models. However, predicted values generally underestimate data, suggesting these models need further tuning to fully reproduce data in this kinematic regime.

Besides heavy hadron production, the LHCb experiment is also suitable to study heavy hadron properties. For instance, LHCb measured the polarisation of $\Lambda$ ($\bar{\Lambda}$) strange (anti)baryons in the $p$Ne fixed-target dataset described before~\cite{LHCb:2024vwi}. Baryons can be produced polarised from strong interaction, with polarisation orthogonal to the production plane given by proton beam and baryon momentum.
Polarisation is measured from the analysis of the $\Lambda\to p\pi^-$ decay distribution
$$p(\cos\theta) = \frac{1}{2} \left(1 + \alpha P \cos\theta \right),$$
in which $\theta$ is proton polar angle in the $\Lambda$ rest frame with respect to the production plane and $\alpha$ is the known decay asymmetry parameter.

The total $\Lambda$ and $\bar{\Lambda}$ polarisations are
$$P_\Lambda [\%] = 2.9 \pm 1.9 (\textrm{stat}) \pm 1.2 (\textrm{syst}),$$
$$P_{\bar{\Lambda}} [\%] = 0.3 \pm 2.3 (\textrm{stat}) \pm 1.4 (\textrm{syst}).$$
Polarisation is also measured differentially as a function of the baryon kinematics. A trend of increasing negative polarisation with Feynman $x_F$ variable is compatible with previous measurements.

\section{Prospects}
The studies presented in these proceedings show the capabilities of the LHCb experiment in analysing heavy-ion and fixed-target collisions collected during the past Run 2 LHC data-taking. At present, the Run 3 data-taking is ongoing with an upgraded LHCb detector, running at higher luminosity and with increased detector efficiency in all data-taking configurations. The upgraded LHCb detector has already collected PbPb and fixed-target datasets, improving its centrality coverage in ion-ion collisions and reaching full coverage in fixed-target Pb-gas events. Moreover, fixed-target datasets recorded with the SMOG2 gas cell have superseded older SMOG ones in terms of integrated luminosity and gas variety.
Meanwhile, an LHCb upgrade phase for future LHC Run 5 data-taking has been proposed~\cite{LHCb:tdr}, including a detailed physics case for heavy-ion and fixed-target physics.

To conclude, a rich physics program with heavy ions is ongoing in full swing at the LHCb experiment, with new and future datasets promising even more valuable results.

\section*{References}


\end{document}